\documentclass[aip, jcp, reprint, groupedaddress]{revtex4-1}

\usepackage{enumitem}
\usepackage{graphicx}          
\usepackage{dcolumn}           
\usepackage{bm}                
\usepackage[mathlines]{lineno} 
\usepackage[utf8]{inputenc}
\usepackage{amsmath}    
\usepackage{amssymb}    
\usepackage{amsthm}
\usepackage{graphicx}   
\usepackage{verbatim}   
\usepackage{color}      
\usepackage{subfigure}  
\usepackage{hyperref}   
\usepackage[capitalise]{cleveref}   
\usepackage{appendix}

\begin{document}

\title{MSM lag time cannot be used for variational model selection}
\author{Brooke E. Husic}

\author{Vijay S. Pande}
\affiliation{Department of Chemistry, Stanford University, Stanford CA 94305, USA}

\date{\today}

\begin{abstract}
The variational principle for conformational dynamics has enabled the systematic construction of Markov state models through the optimization of hyperparameters by approximating the transfer operator.
In this note we discuss why lag time of the operator being approximated must be held constant in the variational approach.
\end{abstract}
\maketitle

Markov state models (MSMs) are a powerful master equation framework for the analysis of molecular dynamics (MD) datasets that involve a complete partition of the conformational space into disjoint states.~\cite{Bowman_Book14} By representing each frame of a MD dataset as its state label, the populations of and conditional pairwise transition probabilities between the states can be counted, leading to thermodynamic and kinetic information about the system, respectively. This information is represented by a transition matrix, which contains all the information necessary to propagate the system forward in time.
The transition matrix is the discrete-time approximation to the transfer operator $\mathcal{T}(\tau)$, which is characterized by its lag time $\tau$. The transfer operator propagates the system, represented by a normalized probability density $u_t(x)$, forward by a time step of $\tau$, and admits a decomposition into eigenfunctions and eigenvalues (see Ref.~\onlinecite{Bowman_Book14}, Ch.~3),

\begin{subequations}
\begin{align} 
&\mathcal{T}(\tau) \circ u_t(x) = u_{t+\tau}(x), \\
&\mathcal{T}(\tau) \circ \psi_i = \lambda_i \psi_i. \label{eq:gep}
\end{align}
\end{subequations}

\noindent{}The eigenvalues $\lambda_i$ are real and numbered in decreasing order. The unique highest eigenvalue $\lambda_1=1$ corresponds to the stationary distribution, and the subsequent eigenvalue/eigenfunction pairs represent dynamical processes in the time series. Importantly, the timescale of each process can only be retrieved with knowledge of the lag time at which the operator was defined using the equation,

\begin{align}\label{eqn:ts}
t_i = -\frac{\tau}{\log \lambda_i}.
\end{align}

Choosing a lag time at which the system is Markovian depends on what type of system is being modeled. At a long enough lag time for the system to be approximated as a Markov process, intrastate transitions occur much more quickly than interstate transitions. The appropriate lag time depends on the system of study:~for protein folding, 50~ns might be appropriate; for electron dynamics, a suitable lag time might be on the order of femtoseconds.
If a system is Markovian at a lag time $\tau$ (if the intrastate transitions occur more quickly than $\tau$), then the system will be Markovian at all lag times greater than $\tau$ and the timescales of the subprocesses will be constant for all Markovian lag times. This idea has motivated the use of implied timescales plots to choose a lag time.~\cite{Swope_JPCB04a} Lag times after which the timescales ``level out'' are assumed to be Markovian, and usually the shortest such time is chosen for the most temporal resolution.

In practice, we usually do not know the true eigenfunctions $\psi$ in~\eqref{eq:gep} and instead need to guess them. For a MSM, this means choosing how to divide phase space into disjoint states. Until recently, choosing how to define the states occupied by a dynamic system represented a bottleneck in MSM methods development, and heuristic, hand-selected states were common. However, the derivation of a variational principle for conformational dynamics by \citet{Noe_MMS13} in 2013 opened the door for a systematic approach to choosing the states of a system. 
Our guess, or \textit{ansatz} eigenfunctions, $\hat{\psi}_i$ will admit corresponding eigenvalues $\hat{\lambda}_i$. Using our \textit{ansatz}, we can state the variational principle derived by No\'{e} and N\"{u}ske,~\cite{Noe_MMS13}

\begin{align}\label{eqn:gmrq}
\text{GMRQ} \equiv \sum_{i=1}^m\hat{\lambda_i} \leq \sum_{i=1}^m\lambda_i,
\end{align}

\noindent{}where GMRQ stands for generalized matrix Rayleigh quotient, which is the form of the approximator when the first $m$ eigenfunctions are estimated simultaneously. By recalling the relation of the eigenvalues and operator lag time $\tau$ to the system timescales in~\eqref{eqn:ts}, we see that the variational principle establishes an upper bound on the timescales of the slowest $m$ processes in the dynamical system. In practical cases, the variational bound can be exceeded due to statistically undersampled processes; therefore, the GMRQ must be evaluated under cross-validation as described in Ref.~\onlinecite{McGibbon_JChemPhys15a}.

\begin{figure}[!h]
    \centering
    \includegraphics[width=0.48\textwidth]{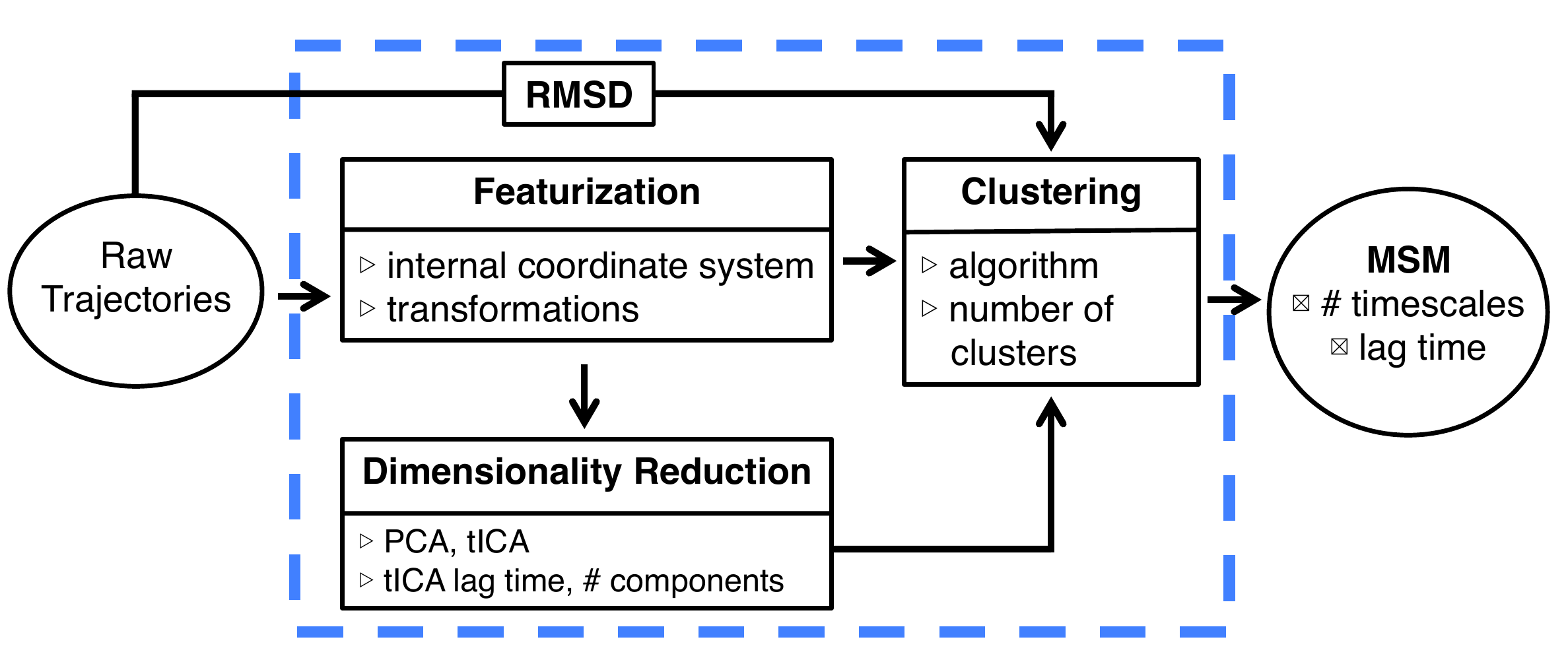}
	\caption{The flow chart shows several ways to create a MSM from raw simulation data. The blue box indicates which of the parameters enumerated can be optimized using the GMRQ. The MSM lag time and number of timescales scored must be held constant. This figure is modified with permission from Ref.~\onlinecite{Husic_JChemPhys16}.}
    \label{fig:chart}
\end{figure}

When we variationally choose a set of eigenfunctions, we can only compare them if we are trying to approximate the same transfer operator. Therefore, the lag time $\tau$ must not be changed when the \textit{ansatz} is changed, and cannot be variationally optimized using the GMRQ---instead, it must be determined using such techniques as implied timescale plots.
In contrast, all transformation and dimensionality reduction choices leading up to the state decomposition are ideal hyperparameters to optimize using the GMRQ. This might include:
\begin{itemize}[itemsep=0.5pt]
\renewcommand\labelitemi{$\checkmark$}
	\item RMSD cutoffs for geometric clustering;
    \item internal coordinate choices such as dihedral angles or contacts pairs, including which angles and pairs to include, and any transformations thereof; 
    \item internal parameters for time-structure based independent component analysis (tICA) such as tICA lag time, number of components retained, and any transformations of these components;
    \item clustering algorithm and number of clusters;
\end{itemize}
\noindent{}but, as discussed above, cannot include:
\begin{itemize}[itemsep=0.5pt]
\renewcommand\labelitemi{$\times$}
	\item the operator lag time, or
    \item the number of timescales scored.
\end{itemize}
\noindent{}These choices are illustrated in Fig.~\ref{fig:chart}. For protein folding, we refer the reader to Ref.~\onlinecite{Husic_JChemPhys16} for a systematic study of these choices in the context of the VAC. 

In practice, we recommend starting with reasonable parameters for the system of study and choosing a valid lag time. Then, at the chosen lag time, perform a hyperparameter search for any of the state decomposition choices listed above. This two step process can then be repeated, alternating lag time validation using fixed hyperparameters with hyperparameter searches for a fixed lag time.

Perhaps the most natural way to understand
the separate treatment of the lag time is to consider its true role in kinetic model building.  While previous MSM approaches have treated it effectively like a hyperparameter (e.g., choosing a lag time based on flattening of implied timescales), in actuality, this approach is fundamentally philosophically incorrect. The lag time must be chosen \textit{a priori} by the researcher, as it directly reflects the resolution of interest to study. Given a method which can directly identify the relevant degrees of freedom, choosing a lag time of picoseconds would bring water dynamics into the state space, vs.~nanoseconds for backbone and side chain dynamics or microseconds for longer time slow scale rearrangements. For this reason, it simply doesn’t make sense to let the model choose the lag time and instead one must have the protocol choose the best model given a pre-chosen set lag time.

MSMs are just one example of the general set of models to which the VAC applies. The popular tICA framework~\cite{PerezHernandez_JChemPhys13, Schwantes_JCTC15} can also be variationally optimized. When using tICA as an intermediate step in MSM construction, the tICA lag time may be varied and optimized. However, in the case where the tICA model is the entity being evaluated, the tICA lag time and number of components scored must be held constant in order to ensure that the same operator is being approximated. Additional extensions of the VAC can be found in Ref.~\onlinecite{Wu_Arxiv17}. We also refer the interested reader to Ref.~\onlinecite{McGibbon_JChemPhys15b}, which presents continuous-time Markov processes that do not have lag times.
We would also like to note that the VAC is not a panacea:~the slowest dynamical processes are often assumed, but not guaranteed, to be the processes of interest, and it is important to verify this for each analysis. We anticipate this note will help guide hyperparameter optimization when using VAC. The open-source software Osprey~\cite{McGibbon_JOSS16} has been designed for this type of optimization and is available on \url{msmbuilder.org}.

The authors are grateful to Matt Harrigan, Carlos Hern{\'a}ndez, Jade Shi, Anton Sinitskiy, Nate Stanely, and Muneeb Sultan for discussion and manuscript feedback.
We acknowledge the National Institutes of Health under No.~NIH R01-GM62868 for funding. V.S.P. is a consultant \& SAB member of Schrodinger, LLC and Globavir, sits on the Board of Directors of Apeel Inc, Freenome Inc, Omada Health, Patient Ping, Rigetti Computing, and is a General Partner at Andreessen Horowitz. 


%

\end{document}